\documentclass[aps,pra,twocolumn,groupedaddress]{revtex4-2}
\usepackage[utf8]{inputenc}
\usepackage{graphicx}
\usepackage{amssymb}
\usepackage{amsmath}
\usepackage[american]{babel}
\usepackage{hyperref}
\usepackage{upgreek}
\usepackage{physics}
\usepackage{pgf}
\usepackage{layouts}

\begin{document}


\title{Limitations of Entangled Two-Photon Absorption detection}

\author{R.~Pollmann}
\affiliation{Integrated Quantum Optics, Department of Physics, Paderborn University, 33098 Paderborn, Germany}
\affiliation{Institute for Photonic Quantum Systems (PhoQS), Paderborn University, 33098 Paderborn, Germany}

\author{F.~Roeder}
\affiliation{Integrated Quantum Optics, Department of Physics, Paderborn University, 33098 Paderborn, Germany}
\affiliation{Institute for Photonic Quantum Systems (PhoQS), Paderborn University, 33098 Paderborn, Germany}

\author{C.~Silberhorn}
\affiliation{Integrated Quantum Optics, Department of Physics, Paderborn University, 33098 Paderborn, Germany}
\affiliation{Institute for Photonic Quantum Systems (PhoQS), Paderborn University, 33098 Paderborn, Germany}

\author{B.~Brecht}
\affiliation{Integrated Quantum Optics, Department of Physics, Paderborn University, 33098 Paderborn, Germany}
\affiliation{Institute for Photonic Quantum Systems (PhoQS), Paderborn University, 33098 Paderborn, Germany}

\date{\today}

\begin{abstract}
    We introduce a method for determining the sensitivity of any given Entangled Two-Photon Absorption (ETPA) measurement.
    By modeling all signal and noise contributions to the measurement, we derive a single numerical value that describes the sensitivity of the ETPA measurement in Göppert-Mayer units.
    This allows us to directly compare vastly different experimental approaches and, determine whether ETPA will be detectable under the given conditions.
    Therefore, we can quantify the effect of any change to a given experimental apparatus and identify the ideal optimization pathway.
\end{abstract}


\maketitle

\section{Introduction}
Two-photon absorption (TPA) is a valuable tool for a variety of spectroscopy and imaging schemes, creating contrast and improving resolution via nonlinear processes~\cite{Mukamel2020}.
However, since any TPA process requires two photons to interact simultaneously with one absorber, the overall interaction cross-sections are prohibitively small.
This limitation is typically overcome by vastly increasing the photon flux with high-power pulsed lasers or high finesse cavities.

Molecular TPA requires the simultaneous arrival of two photons and sets strict limits on their combined energy.
Thus, time-frequency entangled photon pairs promise a significant advantage in absorption yield.
Furthermore, it has been proposed, that the quantum correlations embedded in the photo pair provide access to molecular information inaccessible with classical light sources~\cite{Schlawin2018, Mukamel2020}. 
However, these correlations could just as well decrease the TPA yields in an unfortunate experimental configuration~\cite{fu2023, rodriguezcamargo2025}.

To measure the effect of quantum correlations on TPA yields, the high power laser is replaced by a parametric down conversion (PDC) source that produces time-frequency entangled photon pairs.
However, the increase in absorption efficiency is now counteracted by the decrease in power driving the TPA.
Therefore, observing molecular TPA driven by entangled pairs (ETPA) in the lab is very challenging, primarily due to the prohibitively low overall ETPA rate~\cite{Tabakaev2022, Landes20212, Landes2024, Parzuchowski2021, Parzuchowski2025, Tabakaev2021, Pandya2024, Gaebler2024, He2024}.

There are two approaches to detecting molecular absorption: Measuring attenuation of the transmitted field and measuring fluorescence emitted by the absorber.
In this work, we focus on direct fluorescence detection as described in~\cite{Tabakaev2022, Landes20212, Landes2024, Parzuchowski2021, Parzuchowski2025}.
However, our approach can also be applied to the transmission approach, as presented in~\cite{Gaebler2024, Pandya2024, Freiman2023, yepizgraciano2025}.

Although a fluorescence event is always tied to a molecule reaching an excited state, any information about the excitation pathway is lost.
Here, we will consider three excitation pathways: ETPA, TPA and single-photon or hot band absorption (HBA)~\cite{Mikhaylov2022}.
Therefore, detecting only a fluorescence signal when exciting with PDC light is insufficient for measuring ETPA.
Furthermore, technical imperfections, most notably detector dark counts, significantly contribute to any recorded signal.

Since the overall goal is to investigate the effect of the quantum correlations on the absorption yield, only the strength of the quantum correlations must be varied in the experiment, while all other parameters have to remain constant.
This modulation can be achieved by either separating the correlated photons~\cite{Pandya2024, Tabakaev2021, Corona-Aquino2022} or by introducing losses~\cite{Tabakaev2022, Landes20212, Landes2024, Parzuchowski2021, Parzuchowski2025, Gaebler2024}.

These different methods and the dependence of the ETPA signature on all properties of the excitation beam make comparisons between published (null-) results nontrivial.
Therefore, we developed a unifying framework that allows us to directly compare different ETPA measurements.
By treating the absorber as a black box and focusing on the measurement system, we developed an analytical model that enables us to predict whether observations are possible by calculating the expected signal and noise levels for any given set of experimental parameters.
Furthermore, we can directly calculate the impact of each experimental parameter on the final sensitivity, thus providing a clear path for improvement of any given ETPA experiment.

\section{Model}

Our model is based on calculating signal-to-noise ratios, analogous to the detection threshold derived in~\cite{yepizgraciano2025}.
By modeling both the signal and the noise, and then comparing them, we obtain a lower bound for the detectable ETPA strength in a given experimental configuration.

In the experiment, a total of $N_{rec}$ potential ETPA fluorescence events will be recorded.
As previously mentioned, this will be the sum of all possible excitation pathways
\begin{align}
    N_{rec} = N_{ent} + N_c + N_{h} + N_{dark},\label{eq:contributions}
\end{align}
the desired ETPA contribution $N_{ent}$, an unavoidable classical TPA contribution $N_{c}$ caused by two uncorrelated photons getting absorbed, as well as direct single-photo absorption $N_{h} $~\cite{Mikhaylov2022}.
In the experiment the recorded number of events $N_{rec}$ will also include a non-fluorescence contribution consisting of detector dark counts $N_{dark}$.

Thus, in any ETPA experiment, the amount of correlated pairs interacting with the absorber must be varied while keeping the total number of photons at the absorber - and therefore the resulting non ETPA contribution - constant.
This results in at least two data points: one with maximal correlations of the light field and the largest total fluorescence yield, from now on called signal ($S$) and one background ($B$) measurement recorded with minimal (spoiled) correlations.
An ETPA signature can then be detected if the difference between the two data points is greater than the measurement uncertainty:
\begin{align}
    S - B \geq u (S) + u (B) \label{eq:detection_th}
\end{align}
with $u (S)$ ($u (B)$) being the measurement uncertainty of $S \ (B)$.
Since we assume Poissonian statistics for all measured quantities:
\begin{align}
    u (S) = n_\sigma \sqrt {S} \text{ and }u (B) = n_\sigma \sqrt {B}. \label{eq:def_uncert}
\end{align}
with the $n_\sigma$ accuracy.
Given a sufficient set of experimental parameters, all contributions to $S$ and $B$ are known.
Therefore, equation~\eqref{eq:detection_th} allows us to make quantitative predictions about the detectability of an ETPA signature.

\subsection{Rate estimations\label{sec:rate_estimations}}
To calculate the TPA rates the pair $\phi_{pair}$ and single-photon $\phi_{s,i}$ fluxes at the absorber are needed.
Therefore, we assume an ideal (non-degenerate) PDC source that produces $N_P$ correlated photon pairs per pump pulse.
Since any experiment will involve optical losses between the source and the absorber, which will break the pair correlations, we describe these losses using the transmission coefficients $\eta_{s,i}$ of signal and idler beams respectively.
Using the photon creation $\hat{a}^\dagger_{s,i}$ and annihilation operators $\hat{a}_{s,i}$ for the signal and idler beam at the sample position, the single-photo flux can be written as
\begin{align}
    \phi_{s,i} = \frac{\ev{\hat{a}^\dagger_{s,i} \hat{a}_{s,i}}}{A_{s,i} T_{s,i}} = \frac{\eta_{s,i} N_P}{A_{s,i} T_{s,i}},
\end{align}
with $A_{s,i}$ and $T_{s,i}$ denoting the cross-sectional area and temporal duration of the wave packets at the sample position.
As previously mentioned, the absorber can be excited by two uncorrelated photons~\cite{Landes2024}.
This classical TPA rate can be written as:
\begin{align}
    f_c = \sigma_c N_t \phi_s \phi_i
        = \epsilon_{c} \eta_s \eta_i  N_P^2 \label{eq:Uncoor_abs}.
\end{align}
with the number of illuminated molecules $N_t$, and the classical two-photon absorption coefficient
\begin{align}
    \epsilon_{c} = \frac{N_t \sigma_c}{A^2 T^2}, \label{eq:epsilonC_to_SigmaC}
\end{align}
assuming both beams have the same spatial $A_s = A_i = A$ and temporal $T_s = T_i = T$ shape, ensuring maximal overlap and therefore maximising the TPA yield.

In PDC, the photon pair is defined by the properties of the pump beam and phase patching of the nonlinear medium, resulting in spatial, spectral, and temporal correlations between the signal and idler photons.
We model these quantum correlations by introducing the entanglement area $A_e$ and the entanglement time $T_e$.
These represent the conditioned spatial and temporal distributions of the idler photon, upon the detection of the signal photon~\cite{Parzuchowski2021, Dickinson2025, Pollmann2024, Roeder2024_1, Roeder2024_2}.
Thus, the pair flux at the absorber is given by:
\begin{align}
    \phi_{pair} = \frac{\ev{\hat{a}^\dagger_s \hat{a}^\dagger_i \hat{a}_s \hat{a}_i}}{A A_e T T_e} = \frac{\eta_s \eta_i N_P}{A A_e T T_e}
\end{align}
with the illuminated area $A$, the entanglement Area $A_e$, the duration of the single-photon wavepacket $T$ and the entanglement time of the photon pair $T_e$.
The ETPA rate can then be written as follows:
\begin{align}
    f_{ent}     = \sigma_c N_t \phi_{pair}
    = \epsilon_{e} \eta_s \eta_i N_P \label{eq:ent_fluo}
\end{align}
with the ETPA coefficient
\begin{align}
    \epsilon_{e} = \frac{N_t \sigma_c}{A T A_e T_e} \label{eq:epsilonE_to_SigmaE}
\end{align}
for the specific experimental realization at hand.
Note, that the ETPA rate does not depend on an entangled cross section $\sigma_e$ but only on the classical cross section $\sigma_c$ as the ``nonclassicality'' is fully transferred to the optical field~\cite{Fei1997, Landes2021}.

The quantum enhancement observed in the experiment is now described by
\begin{align}
    \frac{\epsilon_{e}}{\epsilon_{c}} = \frac{A T}{A_e T_e}, \label{eq:quant_adv}
\end{align}
and is completely given by the optical fields, consistent with the arguments made in~\cite{Fei1997, Gaebler2024, Landes2021}.
Interestingly,~\eqref{eq:quant_adv} is the product of the spatial and spectral mode number in the experimentally relevant case of highly multimode PDC. 
Any additional enhancement, as discussed in~\cite{Landes2021, rodriguezcamargo2025}, could be included as an additional factor in~\eqref{eq:quant_adv} without changing the arguments made here.

Even far off resonance, one photon absorption can occur, leading to a linearly scaling fluorescence signal~\cite{Mikhaylov2022}.
This HBA rate can be expressed using the signal (idler) flux $\phi_{s,i}$ at the absorber as,
\begin{align}
    f_{h}     = N_t (\sigma_{h, s} \phi_s + \sigma_{h, i} \phi_i)
                = (\epsilon_{h, s} \eta_s + \epsilon_{h, i} \eta_i) N_P.               \label{eq:HBA}
\end{align}
With the wavelength dependent hot band cross-section $\sigma_{h, s ,i}$ and consequently $\epsilon_{h, s, i}$.
Since all experiments we analyse in this work utilise frequency degenerate photon pairs, $\epsilon_{h, s} = \epsilon_{h, i} = \epsilon_{h}$.

\section{Sensitivity calculation}
\begin{figure}
    \center
    \input{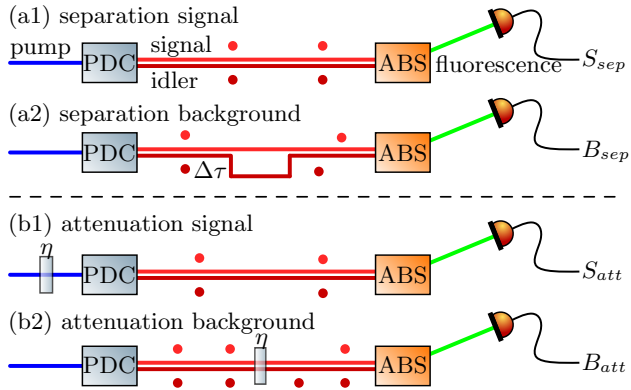}
    \caption{Experimental configuration for Signal and Background measurements in both schemes discussed here.
        Entangled photon pairs are created by pumping a PDC crystal and are sent to a two-photon resonant absorber (ABS).
        The resulting fluorescence is then detected by a single-photon-sensitive detector.
        A setup for measuring the signal $S_{sep}$ in the separation scheme is shown in (a1).
        The background measurement $B_{sep}$ is obtained by delaying the idler beam by $\Delta \tau$ (a2).
        In the attenuation scheme, an attenuator $\eta$ is placed either in the pump beam to obtain the signal measurement $S_{att}$ (b1) or in the signal and idler beams to measure the background $B_{att}$ (b2).
    }\label{fig:setups}
\end{figure}

Since TPA and HBA processes are always present, even for an ideal stream of entangled pairs, and cannot be fully suppressed by clever design, their impact on any ETPA measurement must be considered.
Using eq.~\eqref{eq:detection_th}, we analyse two different measurement schemes (shown in figure~\ref{fig:setups}), by calculating the expected number of detected events and the corresponding uncertainties in an ETPA measurement.

So far, all derivations have been made for a single pump pulse.
Therefore, to find the actual measured values, the repetition rate of the pump $f_{rep}$ and the integration time $T_{int}$ have to be considered.
In the case of a continuous wave pump, we use the coherence time of the pump (1/bandwidth) as the pulse duration $T$ and set $f_{rep} = 1/T$.

\subsection{Separation\label{sec:sepearation}}

In this conceptually straightforward approach, one measurement $S_{sep}$ is taken with pairs that are as strongly correlated in time and space as possible, and the second measurement $(B_{sep})$ is taken with the pairs separated in time (space) by more than the entanglement time (area), breaking all pair correlations.
We assume that the delay matches the time between pump pulses $\Delta\tau = \frac{1}{f_{rep}}$, as to not change the single-photon flux at the absorber.
An experimental realization of the method in the spatial domain can be found in~\cite{Pandya2024}.
Based on the considerations in section~\ref{sec:rate_estimations} and eq~\eqref{eq:contributions} the expected signal and background can be written as follows:
\begin{widetext}
\begin{align}
S_{sep}  =T_{int} f_{rep} T & \biggl(
\underbrace{\eta_s \eta_i \eta_d N_P^2 \epsilon_{c}}_{\text{CTPA}}
+ \underbrace{\eta_s \eta_i \eta_d N_P  \epsilon_{e}}_{\text{ETPA}}  
+  \underbrace{N_P \eta_d (\eta_s + \eta_i)\epsilon_{h}}_{\text{HBA}}
+ \frac{f_{dark}}{T f_{rep} }  \biggr) \label{eq:s_sep} \\
B_{sep}  =T_{int} f_{rep} T & \biggl( \eta_s \eta_i \eta_d N_P^2 \epsilon_{c} 
+ N_P \eta_d (\eta_s + \eta_i)\epsilon_{h}+ \frac{f_{dark}}{T f_{rep} }  \biggr) \label{eq:B_dist}
\end{align}
In addition to integrating over the acquisition time ($T_{int}$) we also included the detection efficiency $\eta_d$ and the detector dark count rate $f_{dark}$.
Note that the ETPA term is missing from~\eqref{eq:B_dist} since signal and idler are separated far enough to break all correlations.
Inserting into~\eqref{eq:detection_th}
and isolating the TPA cross-section $\sigma_c$ by using~\eqref{eq:epsilonC_to_SigmaC} and~\eqref{eq:quant_adv} yields
\begin{align}
\sigma_c \geq & \frac{ n_\sigma ^2 A_e^2 T_e^2}{T_{int} f_{rep} T \eta_s \eta_i \eta_d N_t} 
\left( 2 + \frac{A T}{A_e T_e N_P} + \right. 
\left. 2 \sqrt{1 + \frac{A T}{A_e T_e N_P} + \frac{A^2 T^2 T_{int}}{A_e^2 T_e^2 n_\sigma^2} \left( \frac{f_{rep} T}{N_P} \eta_d (\eta_s + \eta_i) \epsilon_{h} + \frac{f_{dark}}{N_P^2}\right) }
\right).\label{eq:sig_sep_det}
\end{align}
\end{widetext}
As mentioned earlier, successful ETPA detection is only possible if the inequality~\eqref{eq:detection_th} and therefore~\eqref{eq:sig_sep_det} is fulfilled.
Thus, we found a lower bound for the detectable TPA cross-section, and therefore the sensitivity of the measurement.
A quick analysis of~\eqref{eq:sig_sep_det} shows, that increasing $T_{int}$ is beneficial in any case.
While in the limit of infinite flux, $\lim\limits_{N_P\to \infty}$~\eqref{eq:sig_sep_det} converges to
\begin{align}
    \sigma_c \geq \frac{4 n_\sigma ^2 A_e^2 T_e^2}{T_{int} f_{rep} T \eta_s \eta_i \eta_d N_t}. \label{eq:sig_sep_det_lim}
\end{align}
Therefore, increasing the number of photon pairs per pulse $N_P$ will lead to diminishing returns.
A more thorough discussion of the implications for ETPA experiment design can be found in section~\ref{sec:results}.

\subsection{Attenuation}
Next, we apply our model to the most prevalent method in the current literature~\cite{Tabakaev2022, Landes20212, Landes2024, Parzuchowski2021, Parzuchowski2025, He2024, Dickinson2025}.
In this method an attenuator is used to vary the strength of the correlation for the signal and background measurements, while ensuring that the single-photon flux remains constant.
The background $B_{att}$ is recorded by placing an attenuator with transmittance $\eta$ in the signal and idler beams as shown in Fig.~\ref{fig:setups} (b2).
This results in a linear reduction of single photons and a quadratic reduction of entangled pairs.
To record the signal $S_{att}$ the same amount of attenuation is now introduced to the PDC pump (see Fig.~\ref{fig:setups} (b1)), linearly reducing the amount of single photons and entangled pairs.
Thus, the signal and background become:
\begin{widetext}
\begin{align}
S_{att} =& T_{int} f_{rep} T
\left( \eta_s \eta_i \eta_d \eta N_P \epsilon_{e} + \eta_s \eta_i \eta_d \eta^2 N_P^2\epsilon_{c}
+ \eta N_P \eta_d (\eta_s + \eta_i) \epsilon_{h}+ \frac{f_{dark}}{T f_{rep}} \right)\\
B_{att} =& T_{int} f_{rep} T
\left( \eta_s \eta_i \eta_d \eta^2 N_P \epsilon_{e} + \eta_s \eta_i \eta_d \eta^2 N_P^2 \epsilon_{c}
+ \eta N_P \eta_d (\eta_s + \eta_i) \epsilon_{h}+ \frac{f_{dark}}{T f_{rep}} \right)
\end{align}
Inserting into~\eqref{eq:detection_th} and isolating the TPA cross-section $\sigma_c$ using~\eqref{eq:epsilonC_to_SigmaC} and~\eqref{eq:quant_adv} yields:
\begin{align}
&\sigma_c \geq \frac{n_\sigma ^2 A_e^2 T_e^2}{T_{int} f_{rep} T \eta_s \eta_i \eta_d N_t} \frac{1}{(\eta-1)^2}
\left\{ 2 + \frac{\eta+1}{\eta} \frac{A T}{A_e T_e N_P}+ \right. \notag \\
& \left. 2 \sqrt{ 1 + \frac{\eta+1}{\eta} \frac{A T}{A_e T_e N_P} + \frac{1}{\eta} \frac{A^2 T^2}{A_e^2 T_e^2 N_P^2} + \frac{A^2 T^2 T_{int}}{A_e^2 T_e^2 n_\sigma^2}(\eta-1)^2  \left(\frac{f_{rep} T}{\eta N_P} \eta_d(\eta_i+\eta_s) \epsilon_{h} +\frac{f_{dark}}{\eta^2 N_P^2}\right) } \right\}
\label{eq:sig_att}
\end{align}
\end{widetext}
To achieve optimal sensitivity with this method, the optimal attenuation value $\eta_{opt}$ must be found.
In the limit of infinite photon pairs $\lim\limits_{N_P\to \infty}$\eqref{eq:sig_att} converges to $\frac{1}{(\eta-1)^2}\times$\eqref{eq:sig_sep_det_lim}.
Yielding the same sensitivity as the separation method in the rather unphysical case of $\eta_{opt} \to 0$.

While $\eta_{opt}$ can take any value between 0 and 1, the actual value of $\eta_{opt}$ can reveal the dominant noise source in the experiment.
If dark counts dominante, then $\eta_{opt} = \frac{1}{2}$, and if HBA dominates, then $\eta_{opt} = \frac{1}{3}$.
Whereas, if the other terms dominate, then $\eta_{opt} = \frac{1}{4}$ or if the constant terms dominate $\eta_{opt} \to 0$ as discussed above.
Unless stated otherwise, we will always use $\eta_{opt}$ in the following discussion.

\section{Results\label{sec:results}}
\begin{figure}[h]
    \begin{center}
        \resizebox{\columnwidth}{!}{\input{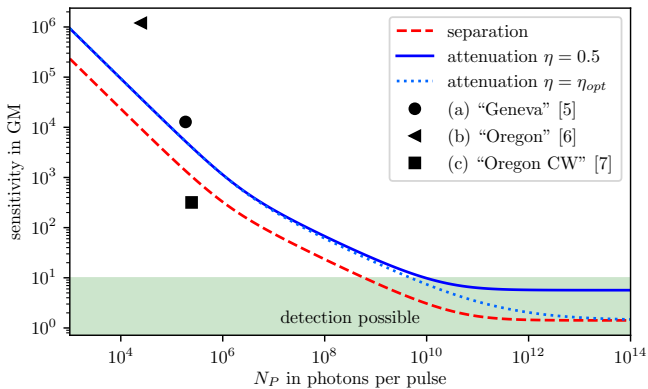}}
    \end{center}
        \caption{Comparing the attenuation method~\eqref{eq:sig_att} (blue) with different attenuation values $\eta$ to the separation method~\eqref{eq:sig_sep_det}.
        The shaded area indicates possible ETPA detection using Rhodamine 6G at 1064\,nm as the target.
        The separation method outperforms the attenuation method up to the case of very high flux, where the two methods converge if $\eta$ is chosen optimally.
        The markers represent the parameter sets used in the attenuation method measurements published in~\cite{Landes20212, Landes2024, Tabakaev2022}.
        The full set of parameters can be found in table~\ref{tab:other_people}.
    }\label{fig:compare_mathodes}
\end{figure}

Figure~\ref{fig:compare_mathodes} shows the minimal detectable value of $\sigma_c$ or ETPA sensitivity for the two methods over the photon pair flux $N_P$. 
In this figure, we assume Rhodamine 6G as the absorber and broadband frequency-degenerate PDC light centred at 1064\,nm.
While implementing the attenuation method is relatively straightforward, deterministically splitting the degenerate beam into signal and idler is more challenging.
One possible approach is to use a dichroic mirror with a transmission edge at the central wavelength to create a short-wavelength signal and a long-wavelength idler.
Then, the idler beam can be delayed by $\Delta \tau$ before recombining the two beams on another dichroic mirror.
Thus, only breaking pair correlations without changing any other property of the exciting beam.
We also applied our method to probabilistic splitting of the degenerate PDC light, see Appendix~\ref{sec:prob_split}.

To compare the different methods we make the strong assumption, that all further experimental parameters are kept constant.
The exact parameters we used can be found in table~\ref{tab:other_people}.

As expected, the separation method outperforms the attenuation method in nearly all cases due to the total suppression of ETPA in the background measurement.

We also plot our predicted ETPA sensitivity for three published experiments using Rhodamine 6G and degenerate PDC centered at 1064\,nm ~\cite{Tabakaev2022, Landes20212, Landes2024} as black markers in figure~\ref{fig:compare_mathodes}.
Since none of the markers lie within the green shaded region, we predict no ETPA detection for these experiments.
However, we would like to stress that our treatment of the absorber as a black box might be insufficient to reproduce the experimental results.
Moreover, it also seems challenging to determine the exact experimental parameters for these types of experiments.
\begin{figure}[ht]
    \begin{center}
        \resizebox{\columnwidth}{!}{\input{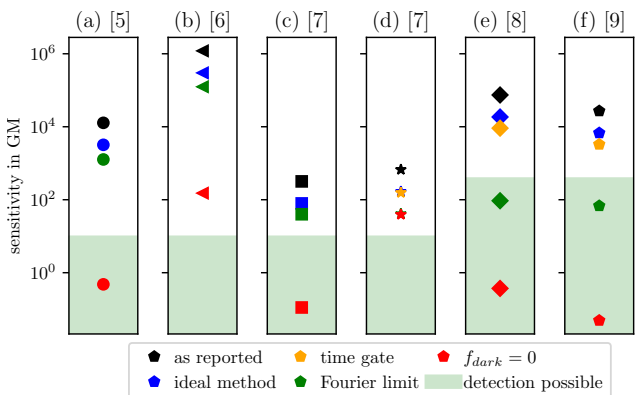}}
    \end{center}
    \caption{Our model applied to published experimental data (black).
        We optimized the sensitivity by selecting the optimal method (blue) and, in the cases of pulsed excitation, introduced time-gated detection (yellow).
        Additionally, we assume Fourier-limited Gaussian photon pairs (green), reaching the sensitivity goal in cases e and f.
        Finally, we eliminate detector dark counts (red) and also reach the goal in cases a and c.
    }\label{fig:no_dark}
\end{figure}

Nevertheless, we can use our model to evaluate possible improvements to future ETPA measurements.
We use the six distinct experimental configurations presented in~\cite{Tabakaev2022, Landes20212, Landes2024, Parzuchowski2021, Parzuchowski2025} as case studies, with all used parameters listed in table~\ref{tab:other_people}.
We plot the ETPA sensitivity according to our model for all six experiments as black markers in figure~\ref{fig:no_dark} and indicate the goal $\sigma_c$ of the particular absorber used as a green shaded area.

First, we determine the ideal method for each experiment under the strong assumption that the experimental parameters are unaffected by the change in method.
While the exact parameters of these experiments vary, they could all improve their sensitivity and reach a lower detectable $\sigma_c$ by using the separation method as indicated by the blue marker.

As the next step we incorporate time-gated fluorescence detection, as described in Appendix~\ref{sec:timegates} for the experiments using pulsed pump beams (shown in yellow in figure~\ref{fig:no_dark}).
For continuous wave pump beams, we introduced a very fast chopper, but  the decrease in effective acquisition time negated any gain due to time-gating.
Additionally, time gates could be a valuable tool for identifying and eliminating non-fluorescence two-photon effects such as second harmonic generation.

Subsequently, we eliminate any group dispersion of the photon pairs and assume Fourier-limited Gaussian pulses, depicted in green.
As experiments e) ``Boulder FS'' and f) ``Boulder Fibre'' suffer from significant group velocity dispersion, this change is especially effective here and is sufficient to reach the desired sensitivity.

Lastly, we completely eliminate detector dark counts by setting $f_{dark} = 0$.
This yields improvements of multiple orders of magnitude, as shown in red, in all but one case (d) ``Oregon Sq'').
Following these steps we reach the desired sensitivity in four out of six cases.

\section{Conclusion}
We present a pragmatic approach to determining the sensitivity of ETPA measurements based on signal-to-noise ratio calculations.
By treating the absorber as a black box and applying a simple probabilistic TPA model, we can calculate the expected signal and noise contributions in a given ETPA experiment.
We derive our ETPA sensitivity from this calculation as the minimal detectable TPA cross-section.

By comparing this value to the TPA cross-section of the absorber in use, we can determine whether the measurement is expected to be successful.
Therefore, our model allows us to directly compare vastly different experiments.
We can directly compare published ETPA results and are able to reproduce most of them.

Exploring different methods shows, that most experiments could be improved by switching to the separation method discussed here and in~\cite{Pandya2024}.
However, since our assumptions on the resulting new experimental setup are quite crude, a more thorough characterization of the resulting experimental setup would be needed to determine the actual enhancement in sensitivity.

\begin{acknowledgments}
We acknowledge financial support from the Federal Ministry of Education and Research (BMBF) via the grant agreement no. 13N16352 (E2TPA)
\end{acknowledgments}

\appendix

\section{Probabilistic separation\label{sec:prob_split}}
If deterministic splitting of the signal and idler beams is not feasible, a modified version of the separation ansatz discussed in section~\ref{sec:sepearation} can be used~\cite{Tabakaev2021}.
Here, the correlations in the background measurement are spoiled by delaying half of both beams by more than the entanglement time $T_e$. 
Again, assuming the delay matches the time between pump pulses $\Delta\tau = \frac{1}{f_{rep}}$ ensures that the single-photon flux at the absorber remains unchanged.

We proceed analogous to the deterministic case as the signal $S_{sep}$~\eqref{eq:s_sep} remains unchanged, while the background $B_{prob}$ is modified, since the ETPA term is not fully suppressed.
\begin{widetext}
\begin{align}
S_{sep}  =T_{int} f_{rep} T &
\left( \eta_s \eta_i \eta_d N_P^2 \epsilon_{c} + \eta_s \eta_i \eta_d N_P  \epsilon_{e}
+ N_P \eta_d (\eta_s + \eta_i)\epsilon_{h} + \frac{f_{dark}}{T f_{rep} }  \right)\\
B_{prob} =T_{int} f_{rep} T &
\left( \eta_s \eta_i \eta_d N_P^2 \epsilon_{c} + \eta_s \eta_i \eta_d \frac{N_P}{2}  \epsilon_{e}
+ N_P \eta_d (\eta_s + \eta_i)\epsilon_{h} + \frac{f_{dark}}{T f_{rep} }  \right)
\end{align}
Again, inserting into~\eqref{eq:detection_th} and isolating the TPA cross-section $\sigma_c$ using~\eqref{eq:epsilonC_to_SigmaC} and~\eqref{eq:quant_adv} yields:
\begin{align}
\sigma_c \geq & \frac{2 n_\sigma ^2 A_e^2 T_e^2}{T_{int} f_{rep} T \eta_s \eta_i \eta_d N_t}
\left(4 +\frac{3 A T}{A_e T_e N_P } + \right. 
\left. 2 \sqrt{ 4 + \frac{6 A T}{A_e T_e N_P} + \frac{2 A^2 T^2}{A_e^2 T_e^2 N_P^2} + \frac{A^2 T^2 T_{int}}{A_e^2 T_e^2 n_\sigma^2} \left(\frac{  f_{rep} T}{N_P} \eta_d(\eta_i+\eta_s)\epsilon_{h} + \frac{f_{dark}}{N_P^2}\right)}\right).
\label{eq:sig_sep_ran}
\end{align}
\end{widetext}
As expected,~\eqref{eq:sig_sep_ran} and~\eqref{eq:sig_sep_det} are very similar, with the deterministic approach outperforming the probabilistic method with equal parameters.
For example, in the limit of infinite photon pairs $\lim\limits_{N_P\to \infty}$~\eqref{eq:sig_sep_ran} converges to $4 \times$\eqref{eq:sig_sep_det_lim}.

\section{Time gating\label{sec:timegates}}
We assume pure exponential decay of the fluorescence rate with the decay time $\tau$ and that the dark counts are distributed uniformly.
Furthermore, we assume $T \ll \tau$ and approximate the excitation pulse as a Dirac delta function.
Thus, time gating linearly decreases dark counts to $f_{dark, gate} = g f_{dark}$ with the relative width of the gate $g$, and nonlinearly decreases fluorescence counts.
Integrating the normalized temporal fluorescence event distribution of one excitation pulse over all subsequent time gates and multiplying by the detection efficiency $\eta_d$ yields the decreased detection efficiency
\begin{align}
    \eta_{d, gate} = \eta_d  \frac{ 1-e^{\frac{-g}{f_{rep} \tau}}}{1 - e^{\frac{-1}{f_{rep} \tau}}}.
\end{align}
If the assumption $T \ll \tau$ does not hold, we numerically integrate the pump pulse shape to determine $\eta_{d, gate}$.
We then use these modified parameters in our model to determine the ideal time gate width and quantify the increase in sensitivity.

\section{Data table}
\begin{table*}
    \centering
    \caption{Experimental parameters used throughout the text if not stated otherwise.}\label{tab:other_people}
    \begin{tabular}{|l|c|c|c|c|c|c|c|}
\hline
  & (a) ``Geneva'' & (b) ``Oregon'' & (c) ``Oregon CW'' & (d) ``Oregon Sq'' & (e) ``Boulder FS'' & (f) ``Boulder Fibre'' & Fig.~\ref{fig:compare_mathodes} \\ \hline
source & \cite{Tabakaev2022} & \cite{Landes20212} & \cite{Landes2024} & \cite{Landes2024} & \cite{Parzuchowski2021} & \cite{Parzuchowski2025} & this work \\ \hline
$T_{int}$ & 5.56\,h & 0.5\,h & 8.33\,h & 96.0\,h & 0.75\,h & 13.0\,h & 3.0\,h \\ \hline
$\eta_{s,i}$ & 71.0\,\% & 70.0\,\% & 85.3\,\% & 85.3\,\% & 76.0\,\% & 50.0\,\% & 70.0\,\% \\ \hline
$\eta_{d}$ & 5.18\,\% & 1.9\,\% & 8.15\,\% & 8.15\,\% & 4.5\,\% & 1.61\,\% & 7.15\,\% \\ \hline
$A$ & 15.9$\,\upmu$m$^2$ & 12.57$\,\upmu$m$^2$ & 15.9$\,\upmu$m$^2$ & 15.9$\,\upmu$m$^2$ & 1256.64$\,\upmu$m$^2$ & 18.0$\,\upmu$m$^2$ & 12.57$\,\upmu$m$^2$ \\ \hline
$T$ & 200.0\,ns & 200.0\,ns & 200.0\,ns & 8.0\,ps & 110.0\,fs & 110.0\,fs & 100.0\,ns \\ \hline
$A_e$ & 15.9$\,\upmu$m$^2$ & 12.57$\,\upmu$m$^2$ & 15.9$\,\upmu$m$^2$ & 15.9$\,\upmu$m$^2$ & 1.7$\,\upmu$m$^2$ & 2.1$\,\upmu$m$^2$ & 12.57$\,\upmu$m$^2$ \\ \hline
$T_e$ & 140.0\,fs & 100.0\,fs & 110.0\,fs & 110.0\,fs & 1620.0\,fs & 1070.0\,fs & 100.0\,fs \\ \hline
$T_{e, min}$ & 56.0\,fs & 42.0\,fs & 56.0\,fs & 56.0\,fs & 17.0\,fs & 22.0\,fs & 100.0\,fs \\ \hline
$N_P$ & $1.9 \cdot 10^{5}$\,PpP & $2.5 \cdot 10^{4}$\,PpP & $2.4 \cdot 10^{5}$\,PpP & $1.2 \cdot 10^{5}$\,PpP & 147.0\,PpP & 2.0\,PpP & $1.0 \cdot 10^{14}$\,PpP \\ \hline
$f_{rep}$ & 5.0\,MHz & 5.0\,MHz & 5.0\,MHz & 5.0\,MHz & 80.0\,MHz & 80.0\,MHz & 10.0\,MHz \\ \hline
$f_{dark}$ & 215\,Hz & 125\,Hz & 3\,Hz & 3\,Hz & 50\,Hz & 100\,Hz & 10\,Hz \\ \hline
$\sigma_{HBA}$ & $4.5 \cdot 10^{-40}$\,cm$^2$ & $4.5 \cdot 10^{-40}$\,cm$^2$ & $4.5 \cdot 10^{-40}$\,cm$^2$ & $4.5 \cdot 10^{-40}$\,cm$^2$ & $1.0 \cdot 10^{-40}$\,cm$^2$ & $1.0 \cdot 10^{-40}$\,cm$^2$ & $1.0 \cdot 10^{-30}$\,cm$^2$ \\ \hline
$N_t$ & $1.6 \cdot 10^{-15}$\,mol & $4.0 \cdot 10^{-16}$\,mol & $1.6 \cdot 10^{-15}$\,mol & $1.6 \cdot 10^{-15}$\,mol & $2.8 \cdot 10^{-12}$\,mol & $1.5 \cdot 10^{-11}$\,mol & $2.0 \cdot 10^{-16}$\,mol \\ \hline
$\eta$ & 50.0\% & 50.0\% & 50.0\% & 25.0\% & 50.0\% & 50.0\% & 1.0\% \\ \hline
$\sigma_c$ att & $1.3 \cdot 10^{4}$\,GM & $1.2 \cdot 10^{6}$\,GM & 318\,GM & 421\,GM & $7.4 \cdot 10^{4}$\,GM & $2.7 \cdot 10^{4}$\,GM & 1\,GM \\ \hline
$\sigma_c$ split & $3.2 \cdot 10^{3}$\,GM & $3.0 \cdot 10^{5}$\,GM & 79\,GM & 167\,GM & $1.9 \cdot 10^{4}$\,GM & $6.8 \cdot 10^{3}$\,GM & 1\,GM \\ \hline
target $\sigma_c$ & 9.9\,GM & 9.9\,GM & 9.9\,GM & 9.9\,GM & 390.0\,GM & 390.0\,GM & 9.9\,GM \\ \hline
\end{tabular}
\end{table*}

\bibliography{sources}

\end{document}